\documentclass{llncs}
\usepackage[pdftex]{graphicx}
\usepackage{flushend}
\usepackage[tight,footnotesize]{subfigure}
\usepackage{listings}

\begin{document}
\author{Nikos Fotiou\inst{1} 
\and Iakovos Pittaras\inst{1}
\and Spiros Chadoulos\inst{1,2} 
\and Vasilios A. Siris\inst{1}
\and George C. Polyzos\inst{1} 
\and Nikolaos Ipiotis\inst{2} 
\and Stratos Keranidis\inst{3}}
\authorrunning{N. Fotiou et al.}
\title{Authentication, Authorization, and Selective Disclosure for IoT data sharing using Verifiable Credentials and Zero-Knowledge Proofs}
\institute{Mobile Multimedia Laboratory, Department of Informatics\\
School of Information Sciences and Technology\\
Athens University of Economics and Business\\
Evelpidon 47A, 113 62 Athens, Greece\\
\email{\{fotiou,pittaras,spiroscha,vsiris,polyzos\}@aueb.gr} \and
Plegma Labs \\
Neratziotissis 115, 15124, Marousi, Athens, Greece\\
\email{\{sc,ni\}@pleg.ma} \and
DomX IoT Technologies \\
Stratigou Sarafi 48E, 55133, Thessaloniki, Greece \\
\email{stratos@domx.io}
}
\maketitle

\abstract{
As IoT becomes omnipresent vast amounts of data are generated, which can be used for building 
innovative applications. However, interoperability issues and security concerns, prevent harvesting 
the full potentials of these data. In this paper 
we consider the use case of data generated by smart buildings. Buildings are becoming ever 
``smarter'' by integrating IoT devices that improve comfort through sensing and automation. 
However, these devices and their data are usually siloed in specific applications or
manufacturers, even though they can be valuable for various interested stakeholders who 
provide different types of ``over the top'' services, e.g., energy management.  Most data 
sharing techniques follow an ``all or nothing'' approach, creating significant security and 
privacy threats, when even partially revealed, privacy-preserving, data subsets can fuel innovative 
applications. With these in mind we develop a platform that enables controlled, privacy-preserving 
sharing of data items. Our system innovates in two directions: Firstly, it provides a framework 
for allowing discovery and selective disclosure of IoT data without violating their integrity. Secondly, 
it provides a user-friendly, intuitive mechanisms allowing efficient, fine-grained access control over 
the shared data. Our solution leverages recent advances in the areas of Self-Sovereign Identities, 
Verifiable Credentials, and Zero-Knowledge Proofs, and it integrates them in a platform that combines the 
industry-standard authorization framework OAuth 2.0 and the Web of Things specifications.        
}

\section{Introduction}

IoT systems generate vast amounts of data nevertheless, their potential is limited by security and privacy concerns, as well as by the lack of interoperability. A striking example is the case of smart buildings. Smart buildings employ a variety of IoT devices that generate data which support various applications, such as energy management, automations, security and safety, etc. These applications are in most cases siloed and the generated data are only used for the specific purposes of each application. Nevertheless, these data can be valuable for a variety of  stakeholders that are able to 
deliver value-added services for other domains. Energy suppliers represent a key stakeholder that can significantly benefit from both energy and non-energy data that can be collected, either directly by smart building systems or even by legacy systems that are integrated with smart IoT equipment. According to an Accenture study~\cite{accenture}, energy utilities will need to master data analytics in the near future, to continue developing valuable, customer-focused products that go far beyond old business models and plain commodity offerings.
Data analytics can benefit energy utilities in multiple ways: a) successful retention of customers through the delivery of innovative personalized services, b) improved customer targeting and segmentation through consumer profiling, c) improved energy market participation through demand forecasting based on machine learning, d) improved energy savings for end users through optimized demand management and many others.

On the other hand, end-users would be interested in securely making a subset of the data generated by their IoT devices available to these 3rd parties, in a stratified manner, to benefit from the added value of the provided services. Nevertheless, several challenges have to be overcome: a) a uniform and standardized way for advertising/discovering, requesting, and transmitting data should be in place, b) sensitive information should be stripped from the shared data without violating data integrity and provenance, c) an efficient, usable mechanism for expressing and enforcing fine grained access control policies should be available, d) data access rights should be expressed in a rich and verifiable manner. In addition to overcoming these challenges, proposed solutions should encourage interoperability and prevent vendor ``lock-in''.
With these in mind we designed, implemented, and evaluated \emph{SelectShare}: a platform for controlled sharing of IoT data,
focusing on smart buildings.

SelectShare is a system that makes available data from IoT systems located in buildings, and facilitates fine-grained, privacy-preserving data access to controlled subsets of these data, while at the same time ensuring data integrity, provenance verification, authenticity, and interoperability with different types of systems. This is achieved by integrating four components. First, an IoT gateway that exposes a data access API by following W3C's Web of Things specifications~\cite{W3cWoT} facilitating data discovery and data interoperability. Second, 
a data transcoder that collects data from  IoT devices, transcodes them into JSON
objects, and signs them using a digital signature scheme that enables selective disclosure of the
claims included in the JSON, providing at the same time Zero-Knowledge Proofs (ZKPs) of their
integrity. Third
an OAuth 2.0~\cite{oauth} based Verifiable Credential (VC)~\cite{ver} issuing mechanism for generating self-contained, fine-grained access tokens. Finally, an HTTP-proxy that acts as a Policy Enforcement Point (PEP), for controlling access to the IoT gateway, as well as for selectively hiding
parts of the responses generated by the gateway. Using this approach, SelectShare achieves
fine-grained access control with minimal overhead and no modification to the IoT devices.

The remainder of this paper is structured as follows. In Section 2, we introduce our
enabling mechanisms and we discuss related work in this area. In Section 3, we detail the 
design of our architecture. In Section 4, we present the implementation and evaluation
of our solution. We conclude our paper in Section 5.

\section{Background and related work}
\subsection{Verifiable Credentials}
A \emph{Verifiable Credential}~(VC)~\cite{ver} allows an \emph{issuer} to assert some \emph{attributes}
about an entity referred to as the VC \emph{subject}. A VC
includes information about the issuer, the subject, the
asserted attributes, as well as possible constraints (e.g., expiration date). Then,
a VC~\emph{holder} (which is usually the same entity as the VC subject) can prove
to a \emph{verifier} that it owns a VC with certain characteristics. This is usually
achieved by including in the VC an identifier (e.g., a public key), owned by the
holder that enables the holder to generate a \emph{proof of possession} (e.g., a
digital signature with the corresponding private key). The VC verification 
process does not require communication with the VC issuer. 
 
The VC data model allows different VC \emph{types}, which define the attributes a VC should include. 
This provides great flexibility, since VC integrators can define their own types
that fit the purposes of their systems. 
Our system uses a new VC type named \emph{CapabilitiesCredential} that ``describes'' which portion of a data item
a user can access. 

\subsection{BBS+ digital signatures}
BBS+ is a digital signature protocol which is used for signing an \emph{array} of
messages. It was first envisioned by \cite{boneh2004short} (from where it takes its name), touched again in \cite{au2006constant}, re-visited in \cite{camenisch2016anonymous} and is currently under standardization~\cite{bbs_rfc_draft}. 
BBS+ provide the ability to sign an array of individual messages, with only a single \emph{constant size} signature. The signature can be validated given the signer's \emph{Public Key}~(PK) and the entire array of signed messages; this is equivalent to validating a ``traditional'' digital signature, if we consider the array of messages as a single compound message. 

BBS+  can be combined with Zero-Knowledge Proofs (ZKP) allowing an entity to selectively 
hide elements of singed array of messages. In particular, \emph{any} entity that knows the signature and the original signed array of messages, can create a proof of knowledge of the signature while selectively disclosing only a sub-array of the signed messages. The proof size will be linear to the number of un-revealed messages. The proof can be validated with only the signer's public key and the array of revealed messages. 

\subsection{Related Work}
Related systems are using ``attribute-based access control'' (ABAC) (e.g., \cite{Goy2022} \cite{Dim2020})
for achieving similar goals. With ABAC, users own a ``token''
that includes their attributes. Then, a ``policy decision point'' (PDP) decides
whether a user can perform a requested operation based on a list of pre-configured
access control policies. Our system follows an alternative approach: our proposed
solution is in essence a ``capabilities-based access control'' system where users
own a token that describes their capabilities. The main advantage of this approach
is that it removes the need for access control lists.
On the other hand, we recognize that ABAC is useful when access control decision involves user
context; in this case the policy decision process should receive as input
attributes related to the context of the user. Our proxy can be easily
extended to include related mechanisms (e.g., the system presented in~\cite{Sch2018}).
 Similarly our proxy can be extended to accommodate aspects such as 
user behavior (e.g., the solution presented in~\cite{Gjo2019}).

Many systems leverage the blockchain technology to achieve similar targets 
(e.g.,~\cite{Sha2022}\cite{Sha2017}). We postulate that blockchain overhead
cannot be tolerated by a system like ours and a trusted proxy that would mediate
the communication between the blockchain and our system would be required: this trusted
proxy would negate any decentralization advantages of the blockchain technology.
It should be highlighted that many VC systems rely on a blockchain to achieve their
security properties. However, VCs in our solution do not need any blockchain-based system. 

Kratos (initially described in~\cite{Sik2020} and then extended in~\cite{Sik2022})
is a system that wants to achieve similar goals as our solution for home IoT
environments, where an IoT device may be owned by multiple users who may define
different access control policies. Our solution considers that each IoT device is
owned by a single entity, hence our approach is simpler. Additionally, Kratos 
proposes its own, specific mechanisms for expressing policies and rights, whereas
our system relies on existing, open standards; hence our solution can be easily integrated
in existing deployments. 

Our solution assumes that IoT devices produce correct data and it does not consider
any countermeasure against malicious IoT device owners. Our solution can be complemented
by existing solutions that incentivize IoT device owner to provide correct data (e.g.,~\cite{Rei2022}).
Finally, in our solution, the used HTTP proxy is trusted to disclose the appropriate information;
other related works rely on cryptographic constructions for not requiring this trust relationship
(e.g., the work in~\cite{Sun2020} relies on ``Attribute-based encryption''). However,
this comes with the cost of additional computational overhead, as well as with the
overhead of managing encryption keys. 

Our solution extends our previous work presented in~\cite{Fot2022}. In SelectShare,
we consider gateways that interconnect IoT devices that
may be owned by different entities. Additionally, SelectShare assumes that the
data generated by the IoT devices has been collected, singed, and stored in a
storage node, prior being requested.
Finally, SelectShare
leverages ZKPs in order to provide even finer-grained access control.

\section{Architecture}
\begin{figure}
  \centering
  \includegraphics[width=0.9\linewidth]{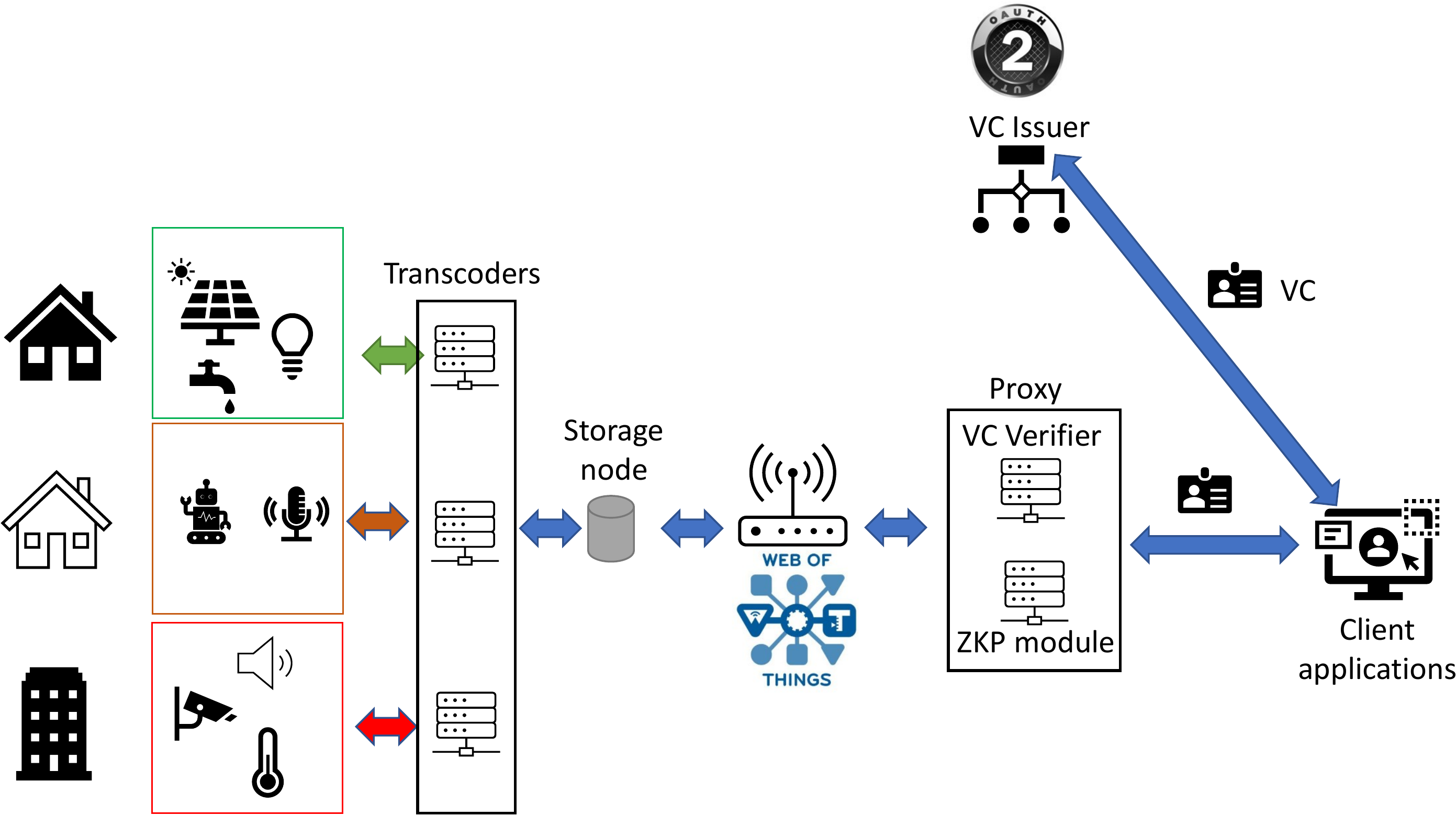}
  \caption {Overview of the SelectShare architecture}
  \label{fig:overview}
\end{figure}

SelectShare architecture (also illustrated in Fig.~\ref{fig:overview}) considers collections of IoT devices the belong to the
corresponding IoT device \emph{owner} (e.g., IoT devices of a smart building).
These devices produce \emph{measurements} that the device owners wish to share with $3^{rd}$
party data \emph{clients} (e.g., analytics services). Data sharing
is implemented through a single gateway, administrated by an independent service
\emph{provider} that can be accessed by clients using a standardized API. 
This gateway retrieves data from a
\emph{storage node}, which acts as a data
repository, populated by specialized data \emph{transcoders}.  
The communication between a client and the gateway is intercepted
by a proxy which is responsible for validating client access rights, as well 
as for hiding parts of the response generated by the gateway.
Clients' access rights are expressed using a Verifiable Credential (VC) issued
by a VC issuer. 

SelectShare considers a setup phase during which: device owners configure
VC issuers with the corresponding access control policies, and the proxy is configured
with a list of trusted issuers per IoT device owner.

\subsection{Data encoding and signing}
In order to facilitate data sharing, SelectShare architecture considers an entity,
named \emph{transcoder}, which transcodes the data produced by each IoT device
based on a predefined JSON \emph{schema}. In our particular instantiation a
generated JSON file includes: i) an IoT device specific identifier and ii) a list
of measurements, where each measurement includes a device-unique measurement identifier (called
\emph{field}) and a list of \emph{value}-\emph{time} pairs. The following listing
is an example of a generated JSON file.
\begin{lstlisting} [label={list:measurement}, caption={A JSON file produced by a transcoder}]
{
  ``deviceID":``monitor-1",
  ``measurements":[
      {
      ``field":``temperature",
      ``values":[
          ``time":``1658162155",
          ``value":``30C"
          ]
      },
      {
      ``field":``humidity",
      ``values":[
          ``time":``1658162155",
          ``value":``50"
        ]
      },
    ]
}
\end{lstlisting} 

It should be highlighted that depending on the requirements of a SelectShare deployment,
different schemas can be considered. A transcoder
is owned and managed by the corresponding IoT devices
owner, i.e., a transcoder interacts with the IoT devices
of a specific owner. Additionally, each transcoder
is configured with a BBS+ signing key and each generated
JSON file is singed using BBS+ (by the transcoder). Finally,
singed JSON files are stored in a storage node, administrated
by the service provider.  

Specific fields of a JSON file can be accessed over HTTP, through SelectShare's
gateway, which implements Web of Things (WoT) Things Description (TD)~\cite{wotTD} specifications.
The WoT architecture attempts to structure well-known web protocols and tools for connecting IoT devices to the Web.  In the WoT architecture communication model, IoT devices (real ones or virtual) are made available through REST-based APIs. 
To improve the interoperability and usability of IoT platforms, the WoT model uses a common format for describing IoT devices referred to as the Thing Description (TD). TD is a JSON-LD encoded file that includes metadata information about the IoT device (such as its id, a title, security definitions, etc), and defines API endpoints that can be used for accessing/invoking a device's properties, actions, and events. 

\subsection{Authentication and Authorization request}
The VC issuer is an OAuth 2.0 authorization server extended with VC issuing capabilities. 
Issued VCs are encoded as JSON Web Tokens (JWT) and signed using JSON Web Signatures (JWS) (based on section 6.3 of~\cite{ver}), improving compatibility and integration with existing tools. 
SelectShare considers VCs that describe which ``measurements'' of the IoT devices of a particular owner,  
a client can access.
These VCs are generated based on policies defined by the
corresponding IoT device owner. 
Additionally, a SelectShare VC issuer maintains a VC revocation list by implementing~\cite{revoc_list}.
In particular, an issuer maintains a revocation list that concerns all non-expired VCs it has issued. This list is a simple bitstring 
and each VC is associated with a position in the list.  Revoking a VC means setting the value of the bit corresponding to the VC 
equal to $1$. Furthermore, each generated VC includes a field named "revocationListIndex" that specifies the position of the credential in the revocation list. 
Finally, a VC issuer is configured with client \emph{credentials} (a client identifier and a client secret
in our implementation), as well as with access control policies that map a client
identifier to the corresponding access rights. 

A client requests from the issuer a VC. A VC request is in essence an OAuth 2.0 access token request using the client credentials grant (section 4.4 of~\cite{oauth}),
(in our implementation the corresponding client identifier and secret are used as the ``credentials grant''). Additionally, the client generates a public-private 
key pair and instructs the issuer to include the generated public key in the issued VC. This is achieved using OAuth 2.0 Rich Authorization Requests~\cite{oauth-rar}. 
In particular, the corresponding OAuth 2.0 access token request, 
is extended to include the generated public key (encoded as a JSON Web Key (JWK)~\cite{rfc7517}) and a digital signature generated using the corresponding private key. 
The issuer authenticates the client based on the included grant and generates a VC. 

A VC is the base64url encoding of a JWT singed by the issuer, according to the VC data model. The generated JWT includes a \emph{cnf} field, as specified by RFC 7800, that contains the public key generated by the client
and included in the corresponding request. 
The VCs used in SelectShare are of type ``CapabilitiesCredential". This type includes an array, called ``capabilities", and each element of this array is a map that maps an IoT device identifier to a list of measurements the client can access.  An example of a VC before encoding follows (the signature part is omitted).

\begin{lstlisting} [label={list:vc}, caption={An example of a VC in our system}]
    {
        ``jti": ``https://issuer.com/credentials/1",
        ``iss": ``https://issuer.com",
        ``aud": ``owner-1''
        ``iat": 1617559370,
        ``exp": 1618423370,
        ``cnf": {
          ``jwk": <client jwk>
         },
        ``vc": {
          ``@context": [
            ``https://www.w3.org/2018/credentials/v1",
            ``https://mm.aueb.gr/contexts/capabilities/v1",
          ],
         ``type": [``VerifiableCredential"],
          ``credentialSubject": {
            ``type": [``CapabilitiesCredential"],
            ``capabilities": {
              ``monitor-1": [
                ``temperature",
              ]
            }
          }
        }
      }      
\end{lstlisting} 

As it can be observed, a VC includes an identifier (the \emph{jti} field),
an identifier for the issuer (the \emph{iss} field),
an identifier for the IoT device owner (the \emph{aud} field), an issuance and expiration time, the client
public key, and the client's ``capabilities''. In the
VC included in this example, a client can access
the ``temperature'' measurement of ``monitor-1'' IoT
device, owned by ``owner-1''.

\subsection{Data access request}
A client application requests to access some measurements of an IoT device by sending
an appropriate HTTP request. This request includes the device identifier as a query
parameter and a list of requested ``fields'' in a HTTP POST body.
The HTTP request includes two HTTP headers: one that contains the JWT-encoded VC, and another
that contains \emph{a proof-of-possession} of the public key included in the VC; the latter proof is generated using OAuth 2.0 Demonstrating Proof-of-Possession at the Application Layer (DPoP)~\cite{dpop}. 
A DPoP proof is essence a JSON Web Signature (JWS) that can be verified using the 
public key included in the corresponding VC. The payload of the JWS is constructed
using a random nonce, the HTTP request method, the HTTP request URI, and a
timestamp indicating the proof's creation time. 

A data access request is intercepted by SelectShare's HTTP proxy.
SelectShare's HTTP proxy includes a \emph{VC verifier}: the VC 
verifier examines if the request includes an appropriate VC and then it 
verifies the validity, the status, and the ownership of a VC. A VC is appropriate
if the ``aud'' claim includes the identifier of the device owner and if contains
the ``fields'' of the ``deviceID'' included in the request.

The validity of a VC is verified by evaluating whether: a) the VC has not expired,
b) the signature of the VC is valid, c) the VC has been issued by an issuer trusted
by the device owner.

The status of the VC is verified by communicating with the VC issuer, and using the validation process described in~\cite{revoc_list}. I.e., in a nutshell, the verifier retrieves the revocation
list (which is a bitstring), locates the bit that corresponds to evaluated VC, and
examines the value of that bit.  

Finally, the ownership of a VC is validated using the DPoP proof, i.e., 
the verifier verifies  that the proof is adequately fresh, it includes a nonce not seen before, 
it includes the correct HTTP method and URI, and its signature can be verified using
the public key included in the VC. 
\subsection{Data access response}
Upon receiving an authorized request, the proxy forwards to the gateway. The gateway
retrieves from a storage node a JSON file that includes \emph{among other things}
the requested fields, and forwards to the proxy. Finally, the proxy applies
the \emph{selective disclosure} process, in order to hide the fields not included
in the client request. The selective disclosure process involves two algorithms: 
\emph{framing} and \emph{canonicalization}.

\subsubsection{Framing}
Framing refers to the derivation of a ``sub-item'' from an item, that contains only part of the original one. Data framing is used to enable selective disclosure of the data item's information. More specifically, the framing algorithm accepts the original item and a frame as input and returns a new item that only contains the key-value pairs specified by the frame. The frame itself is a JSON structure that specifies the parts of the original item that should appear in the resulting one (and be disclosed in the end). For this purpose, the frame contains the keys  that lead to the values that the prover will want to reveal.
The framing algorithm used in SelectShare also includes special symbols that can be used for 
selecting specific elements in an array. For example, considering Listing~\ref{list:measurement} the following frame will reveal ``the value of all measurements that include the field \emph{temperature}":

\begin{lstlisting} [label={list:frame}, caption={An example of frame used in SelectShare}]
  {
    ``measurements": {
        ``*":{
            ``field":``temperature'',
            ``values":{
                ``*":{
                    ``value":``"
                }
            }
        }
    }
}
\end{lstlisting}

Applying this frame in Listing~\ref{list:measurement} will result in the following object:

\begin{lstlisting} [label={list:frameoutput}, caption={Output of framing operation}]
  {
    ``measurements":[
        {
        ``field":``temperature",
        ``values":[
            ``value":``30C"
            ]
        }
      ]
  }
\end{lstlisting}

\subsubsection{Canonicalization}
As discussed previously, BBS+ signatures act on arrays of messages and not on structured data formats like JSON.  In order for a transcoder to be able to sign a data item, as well as in order for the proxy to be able to derive ZKPs, data items must be canonicalized. Various canonicalization algorithms have been proposed by related efforts. A canonicalization algorithm serializes a JSON-encoded item into an array of messages, which can then be signed by a multi-message digital signature system like BBS+. 
There are various security requirements that those algorithms must be conformant with, in order to not compromise the security of the system. In this work, we are using the JCan algorithm~\cite{Kal2022} which is a lightweight, provably secure, JSON canonicalization proposal, designed to work with any data model. 

\subsubsection{Selective disclosure}
Any entity can generate a sub-item of a content item based on a frame and provide a ZKP that proves its correctness as follows. Initially, that entity applies the framing algorithm to derive the sub-item. After framing, the same entity canonicalizes the resulting sub-item, gets the array of messages that correspond to the revealed information (from the security properties of the canonicalization algorithm, this array is guaranteed to be a subset of the signed array that resulted from the canonicalization of the original item) and uses that array to derive a ZKP using BBS+.  

The function of selective disclosure is implemented in a distributed manner by the transcoder and the ZKP module of the proxy.
In particular, transcoders are responsible for signing the generated JSON objects using BBS+ signatures. The signed object is forwarded through the WoT gateway to the proxy. Then the ZKP module of the proxy is responsible for framing the signed object and for generating the corresponding ZKP. The framing operation is implemented by taking into consideration the requested ``fields'' option included. It should be highlighted that the proxy assumes that the user is authorized to access this field: this is true since if the user was not authorized, the incoming request would have been blocked during the VC verification process. 

\section{Implementation and evaluation}
We have implemented SelectShare's issuer\footnote{\url{https://github.com/mmlab-aueb/vc-issuer}} as .net 
core web application. Similarly we have implemented SelectShare's HTTP proxy as 
a Python 3 application\footnote{\url{https://github.com/mmlab-aueb/py-verifier}}. 
Finally, we implemented SelectShare's gateway based on Eclipse's Thingweb WoT
gateway\footnote{\url{https://projects.eclipse.org/projects/iot.thingweb}}.

SelectShare introduces minimal communication overhead. VCs can be long-term (since
they are bound to a public key owned by the client), hence client authorization does
not have to take place often. Similarly, by using DPoP, a client can prove possession
of its VC in a single message, i.e., there is no need for additional roundtrips. 
Moreover, the size of a VC and the corresponding proof is only few bytes. Finally,
when it comes to VC status verification, a VC verifier can retrieve the revocation
list once and use it for multiple requests. It is reminded that a revocation list 
is a bitstring that includes the status of non-expired VCs: since each VC corresponds
to single bit, a revocation list may include thousands of VCs.

\begin{figure}
    \centering
    \includegraphics[width=0.9\linewidth]{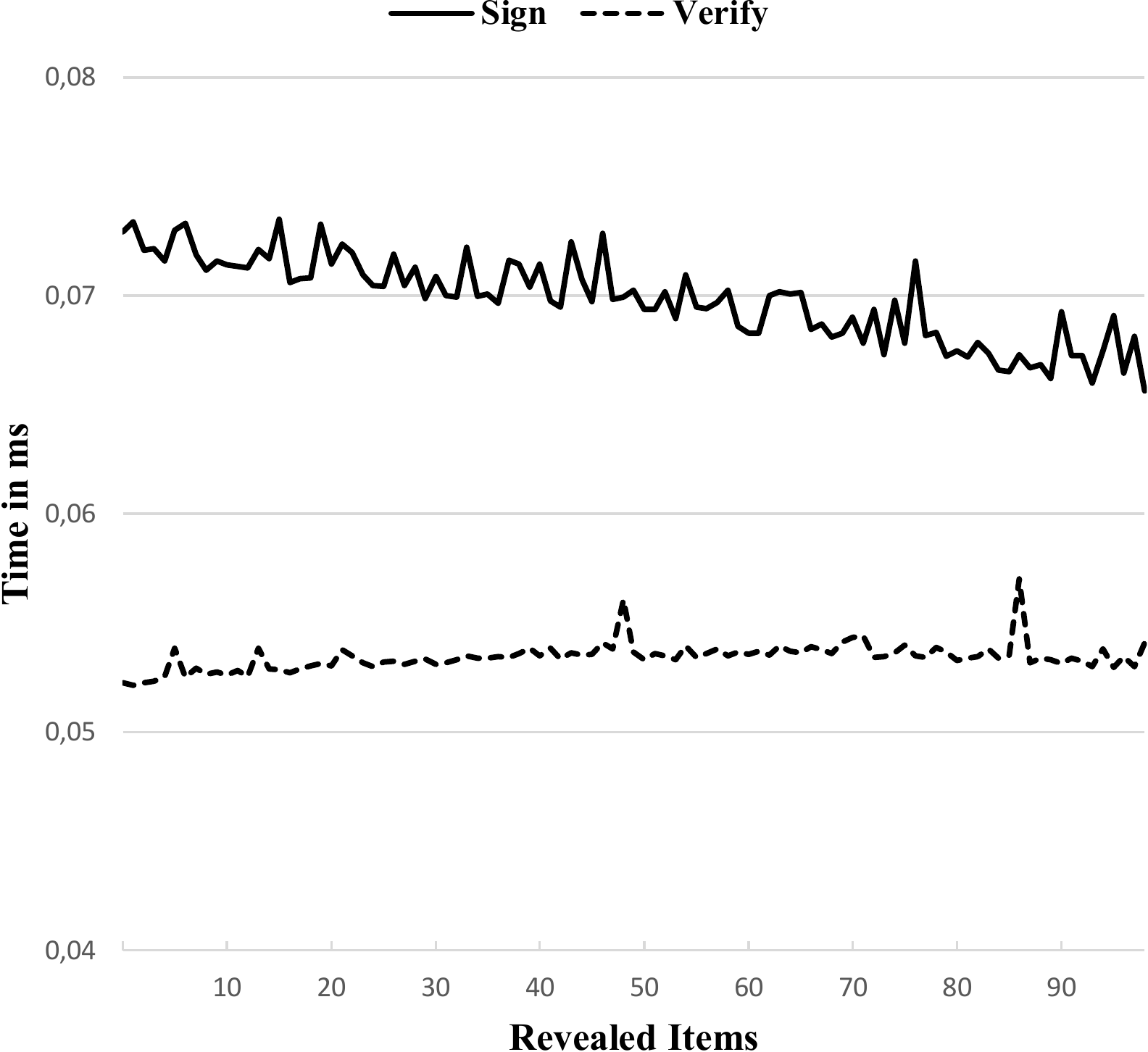}
    \caption {Time to calculate a ZKP as a function of the revealed items}
    \label{fig:over1}
\end{figure}

Similarly SelectShare introduces minimal computational overhead. VC verification process involves only the validation of two digital signatures
as well as a lookup in a JSON object. Both operations are lightweight. 
When it comes to the overhead introduced to a proxy by the selective disclosure process
we performed the following experiment in an Ubuntu 18.04
machine equipped with an intel i7-3770 CPU, 3.40GHz and 16GB of RAM. We
constructed JSON measurement file  composed of 100 fields each of which includes
a single value. We calculate the
time required to sign and verify sub-items that include form $1$ to $99$ values.
Fig.~\ref{fig:over1} show the signature and verification time, measured in ms. It
can be observed that as the number of items included in the sub-item increases, the
signature creation time decreases. This happens because for each hidden item a
proxy has to perform a number  of multiplications. On the other hand, the
signature  verification time remains almost stable. It should be noted that these
measurements are obtained without any ``pre-calculation'', however, in a real deployment
a proxy can pre-calculate many of the computations required to create a ZKP.

\subsection{Security properties}
SelectShare considers the following trust relationships. An IoT device owner trusts: 
the VC issuer to issue an appropriate VC and correctly maintain the revocation list, 
the VC verifier module of the 
proxy to validate VCs and a proofs correctly, and the ZKP module of the proxy to not
reveal ``extra fields''. A client trusts: the VC issuer to correctly maintain the
revocation list, and the proxy to not perform ``denial of service''.  

SelectShare facilitates security management and decreases attacks' surface. In particular,
in SelectShare all access control policies are managed in a single point: the VC issuer.
Adding, updating, or removing an access control policy involves no communication with
the verifiers or the gateway. This is achieved by adopting the ``capabilities-based
access control'' paradigm that removes the need for maintaining access control
lists (as opposed for example to Role/Attribute-based access control). Similarly,
the access control decision process is simple and the most ``advanced'' and error
prone task is examining if the requested resources are included in the provided VCs. 
Related to that, by adopting the JWT encoding for the VCs and by relying on existing
standards, our solution can leverage a plethora of existing tools that perform most
of the tasks required by the access control decision process.  

By adopting the ZKP-based approach for implementing selective disclosure, SelectShare
provides fine-grained access control, preserving at the same time the context of the
output data. For example, in a solution where each ``field'' in a JSON file is 
individually signed, additional measures must be considered in order to prevent
a proxy from creating fake items by ``combining'' fields from different files. 

\section{Conclusions}
In this paper we presented the design and implementation of SelectShare: an access
control solution that allows fine-grained access control for IoT data sharing. SelectShare's
core components are built by leveraging already standardized solutions, which facilitates its
integration with existing systems. Additionally, many security mechanisms of SelectShare
are implemented in a HTTP proxy, hence, existing HTTP-based resources can be
transparently protected. SelectShare facilitates interoperability and improves security
management.

Ongoing and future work in this area includes tighter integration of VC with
the WoT gateway, e.g., the Thing Description generated by the WoT gateway can include
``specifications'' of the expected VCs. Additionally, our system can be extended to 
support other means of client authentication (most notably \emph{Decentralized Identifiers}),
selective disclosure of VCs (achieving this way the principle of least privilege), 
as well as support for advanced trust relationships (e.g., delegation of access rights).

\section*{Acknowledgments}
The work reported in this paper has been funded in part by
European Union's Horizon 2020 research and innovation programme
through subgrant \textit{Selective IoT data sharing (SelectShare)}
of project \textit{NGI DAPSI}.

\bibliographystyle{spmpsci} 
\bibliography{references}

\end{document}